\documentclass[twocolumn,showpacs,preprintnumbers,amsmath,amssymb]{revtex4}

\usepackage{graphicx}
\usepackage{dcolumn}
\usepackage{bm}

\begin{document}

\preprint{APS/123-QED}

\title{Observation of prolonged coherence time of the collective spin wave of atomic ensemble
in a paraffin coated ${}^{87}$Rb vapor cell}

\author{Shuo Jiang${}^1$, Xiao-Ming Luo${}^1$, Li-Qing Chen${}^1$, Po Ning${}^1$, Shuai Chen${^3}$, Jing-Yang Wang${}^2$, Zhi-Ping Zhong${}^2$ and Jian-Wei Pan${}^{1,3}$}
\affiliation{$^{1}$Physics Department, Tsinghua University,
Beijing 100084, P.R.China\\ $^{2}$College of Physical Sciences, Graduate University
of the Chinese Academy of Sciences, PO Box 4588, Beijing 100049\\
$^{3}$Physikalisches Institute, Universit\"at Heidelberg, 69120
Heidelberg, Germany
}


\begin{abstract}
We report a prolonged coherence time of the collective spin wave of
a thermal ${}^{87}$Rb atomic ensemble in a paraffin coated cell. The
spin wave is prepared through a stimulated Raman Process. The long
coherence time time is achieved by prolonging the lifetime of the
spins with paraffin coating and minimize dephasing with optimal
experimental configuration. The observation of the long time
delayed-stimulated Stokes signal in the writing process suggests the
prolonged lifetime of the prepared spins; a direct measurement of
the decay of anti-Stokes signal in the reading process shows the
coherence time is up to 300$\mu$s after minimizing dephasing. This
is one hundred times longer than the reported coherence time in the
similar experiments in thermal atomic ensembles based on the
Duan-Lukin-Cirac-Zoller (DLCZ) and its improved protocols. This
prolonged coherence time sets the upper limit of the memory time in
quantum repeaters based on such protocols, which is crucial for the
realization of long-distance quantum communication. The previous
reported fluorescence background in the writing process due to
collision in a sample cell with buffer gas is also reduced in a cell
without buffer gas.
\end{abstract}

\maketitle Quantum communication is the absolute secure method for
the information transfer. At present the communication distance is
limited to hundred-kilometer scale mainly due to the inevitable loss
of the photon during the transfer. To tackle this problem, a model
of quantum repeater~\cite{1} is introduced, combining the quantum
memory, entanglement purification and entanglement swapping. In the
protocols to realize it, such as the DLCZ protocol~\cite{2} and the
following improved schemes~\cite{3,4,5}, quantum memory is essential
to increase the success probability of such protocols because the
generation of entanglement states are probabilistic.
\par
The quantum memory is achieved by coherent manipulation of the
atomic states, such as preparing a total symmetric collective
state (spin wave) by Raman scattering process~\cite{2}. The qubit
can then be stored and retrieved by manipulating such state. The
coherence time of this collective state will determine the memory
time of the quantum repeater. Long memory time is desirable for
the long-distance quantum communication. For instance, to
establish entanglement of two qubits over hundred-kilometer scale,
the memory time needs to be on the order of hundred-microsecond.
\par
Several groups are presently working on the implementation of the
quantum repeater based on the DLCZ and its improved protocols and
have achieved many significant experimental
advances~\cite{6,7,8,9,10,11,12,13,14,15,16,17}. This includes the
coherent manipulation of the atomic states through Raman process in
an atomic ensemble~\cite{6,7,8,9,10,11}; entanglement of the atomic
states~\cite{12,13,14,15} and the realization of the building block
of the quantum repeater~\cite{16,17}. Both the thermal vapor and
cold atomic ensembles are employed in those experiments. Most
reported coherence time based on cold atoms is in the time scale of
10 microseconds~\cite{10,13,16}, which was attributed mainly limited
by the residual magnetic field. For thermal atoms, the bottleneck of
the coherence time was reported to be the atomic diffusion out of
the read laser region and the inelastic collisions between the atoms
and glass wall, the reported cohererence time was only about
3$\mu$s~\cite{8}.
\par
During the preparation of this paper, a study on cold atomic
ensemble which is simultaneously carried out by another group also
led by Jian-Wei Pan at Heidelberg recently reported ms coherence
time in cold Rb ensemble~\cite{18}. This is achieved by reducing the
effect of magnetic field using 'clock state' and the dephasing
caused by random atomic motion with correct detection configuration.
\par
The thermal atomic vapor cells enjoy simple technique and better
magnetic shielding, but in regular glass cells, the coherence time
of the prepared spin wave is limited by the collisions between walls
and atoms themselves. In the regular glass cell collisions with wall
will cause the spin flip, reducing its lifetime. The time scale is
on the order of tens us for a realistic pencil shaped cell (cm in
dimension), which is much faster than the collision between atoms (
under our temperature T = 78 $^{o}$C, the atomic density is
$10^{12}$/$cm^{3}$, and the mean collision time is on the order of
ms). A cell with paraffin coated walls~\cite{19} would exceed this
limit, since paraffin coating had been demonstrated in reducing such
destructive collisions, allowing atoms to undergo many elastic
wall-collisions and prolong the spin lifetime up to 1
second~\cite{20}. The paraffin-coated alkali-vapor cells had been
successfully used in entanglement between atomic ensembles with
continuous variables~\cite{21}, and slow light experiment~\cite{22}.
To our knowledge, the study on the coherence time of spin waves in a
paraffin coated cell based on the DLCZ protocol and the effect of
dephasing as reported by Zhao \textit{et al.}~\cite{18} in a thermal
ensemble has not been reported yet. In order to prolong the
coherence time, buffer gas is often filled in the sample cell, it
would slow the diffusion speed of atoms to some extent, but also
brought a new problem~\cite{23}: an additional and considerable
fluorescence noise signal, caused by collisional perturbation of the
excited state due to buffer gas during the write (read) process,
severely limited the fidelity of quantum communication.
\par
In this letter, we investigate the capability of applying the
thermal atomic ensembles as candidate for quantum repeater, by
studying the coherence time of the spin wave using $^{87}$Rb vapor
in a buffer-gas-free paraffin coated cell (P-cell). Experiments were
carried out to test whether the prolonged lifetime of the spins
under paraffin coating and minimizing the dephasing process could
enable long coherence time in a thermal sample as well. This
coherence time will provide the upper limit of the memory time in
the quantum repeater. We also investigate whether the
buffer-gas-free environment would reduce the collision-induced
fluorescence background. Though in the DLCZ protocol, the coherent
manipulation of atomic state is implemented by spontaneous Raman
scattering with single excitation, stimulated Raman scattering could
also be used to realize such goal, as Raymer group~\cite{24} pointed
out. The coherence time would be same for both cases, and we measure
it in a P-cell in the stimulated Raman region for the ease of signal
detection and background distinction.
\par
\begin{figure}
\includegraphics[width=8cm,height=6cm]{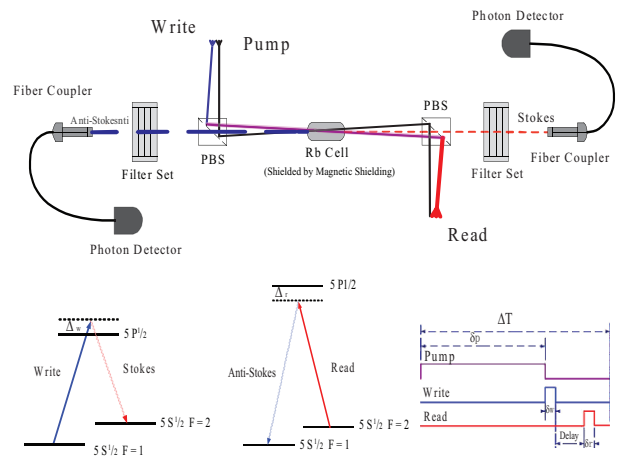}
\caption{Schematic of experiment for generation of atomic
collective state. Off-axis, counter-propagating geometry of Braje
et al.[22] is adopted in our experiment, in which write and read
pulses collinearly propagate into a Rb cell in sequence and we
collect the generated output fields (Stokes, Anti-Stokes), the
direction of this collection forms a 2$^{0}$ angle with the
direction of the write-read input light. PBS stands for polarizing
beam splitters; the Filter Set is composed of selective absorption
by atomic cells and narrow-band filter. The inset illustrates the
relevant atomic level scheme. $\Delta_w = $1GHz, $\Delta_r =
$400MHz} \label{fig:fig1}
\end{figure}

Figure~\ref{fig:fig1} provides an overview of our experimental
setup. The $^{87}$Rb atoms are in a P-cell which is 50mm(length) by
5mm(diameter) and is heated to 78$^{o}C$. The sample cell is put in
a 3-layer magnetic shielding with residual field inside ~10nT. A
pump laser in resonant with $|5S_{1/2}, F=2\rangle$ to
$|5P_{1/2}\rangle$ is first turned on for 50$\mu$s to pump atoms to
$|5S_{1/2},F=1\rangle$; then the write laser whose frequency is blue
shifted 1GHz with respect to the $|5S_{1/2},F=1\rangle$ to
$|5P_{1/2},F'=2\rangle$ is turned on whose duration depends on the
laser intensity, lasted till the stimulated Stokes signal is
observed; after a controlable delay the read laser pulse of 2us
whose frequency is red shifted 400MHz with respect to the
$|5S_{1/2}, F=2\rangle$ to $|5P_{1/2},F'=1\rangle$ is turned on. The
write-read beams are counter propagating and the collection of
Stokes and Anti-Stokes signals by optical fibers is also collinear.
Initially we chose a skewed configuration as Braje \textit{et
al.}~\cite{23} where the two directions forms an angle $\theta =
2^{o}$ to minimized the background laser signal. The signals, after
the polarization filter and frequency filter, were passed through a
scanning Fabry-Perot (F-P) spectrometer (Finesse 200, 7.5GHz FSR,
transmission 10\%) and the output was fed to the single photon
detector. Thus allow us to analyze the intensity of the frequency
components and single out the Stokes and anti-Stokes signals. The
write laser has a maximum power of 7mW and read laser of 80mW at the
input face of the cell. The lasers are locked at the respective fine
structures of the $^{87}$Rb absorption peaks, and the frequency
detuning and pulse generation are achieved by AOMs. The detection of
the Stokes signal signifies the preparation of the spin wave, and
the decay of the anti-Stokes signal at specified direction in the
reading process at different time delay is used to measure the
coherence time of the spin wave. This is essentially a Time-resolved
Coherent Anti-stokes Raman Scattering (T-CARS) method.
\par
\begin{figure}
\includegraphics[width=8.6cm,height=3.1cm]{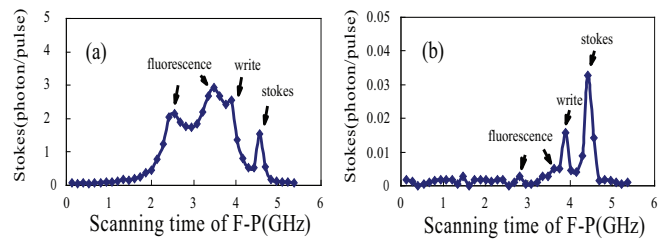}
\caption{Stimulated Stokes signal after a F-P. (a) is the result
in a glass cell with 7 torr Neon buffer gas; (b) is the result in
a P-cell with no buffer gas.  The pulse width of write is 2
$\mu$s. Only polarization and narrow-band filter is applied to let
the signal of the write laser appear as wavelength marker. With
further frequency filter by atomic cells, the write light can be
completely suppressed.} \label{fig:fig2}
\end{figure}
First we report the frequency analyze of the signals during the
writing process in the buffer-gas-free P-Cell. Fig~\ref{fig:fig2}(a)
is from the regular cell with 7 torr neon buffer gas where a
considerable fluorescence signal is observed besides the Stokes;
Fig~\ref{fig:fig2}(b) is from a buffer-gas-free P-cell. It is clear
that the Stokes signal in this P-cell is much purer in a sense that
the fluorescence background is greatly reduced. Thus the problem of
collision-induced fluorescence background as reported by Manz
\textit{et al.}~\cite{23} in the buffer gas cell is suppressed by
using buffer-gas-free sample cells.
\par
\begin{figure}
\includegraphics[width=8.6cm,height=3.1cm]{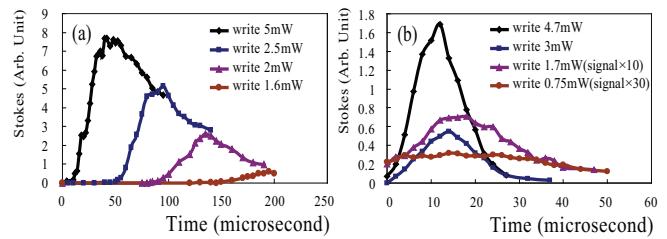}
\caption{(a)The long-delayed stimulated Raman Stokes signal in the
paraffin-coated Rb cell. The write laser has a waist of 3mm inside
the cell and its power is varied. Time zero is the turn-on of the
write pulse which lasts for 200$\mu$s. (b) In the non-coated Rb
cell, as the intensity of the write light decreases, the
stimulated excitation transformed to spontaneous excitation
without any long-delayed stimulated excitation observed. The long
delayed rise of stimulated Stokes in P-cell suggests the paraffin
coating helps in protecting the Rb spin from being destroyed by
the collision with the wall.} \label{fig:fig3}
\end{figure}

Next we measure the temporal pulse shape of the Stokes signal in the
writing process. This is achieved by using a long write pulse and
monitoring the Stokes signal after the F-P, set the gate (1$\mu$s
long) of single photon detector at various temporal positions of the
write pulse, the results are shown in Fig~\ref{fig:fig3}. An
interesting phenomenon is observed for the first time, we discover a
long-delayed stimulated Stokes at various writing powers. A Stokes
signal is generated by spontaneous Raman scattering during the first
microseconds after the write light is turned on, corresponding to
the flat region at early time in Fig~\ref{fig:fig3}(a), and
increased exponentially at later time to reach the stimulated Raman
scattering. Fig~\ref{fig:fig3}(a) shows the variations of this rise
time at different write powers. Depending on the power, the rise
time of the simulated Stokes can be as long as 150 $\mu$s.
\par
Phenomenon of this long-delayed stimulated Raman in P- cell is in
great contrast to what we observed in a non-coated glass cell in
Fig~\ref{fig:fig3}(b). Here the rise of the stimulated Stokes is
only on the order of a few $\mu$s. The rise time is delayed at lower
laser power but only about 10$\mu$s, further reduce the power would
make the stimulated signal disappear. This phenomenon is the result
of longer lifetime of the spins in a P-cell and can be qualitatively
explained as follows: The stimulated emission happens only when the
number of excited spins or the amplitude of the spin wave exceed
certain threshold. The long time delayed stimulated scattering in
the P-cell could only happen if there is an accumulation of the spin
wave, i.e. the decay of the spin wave is slower than its generation
by the write pulse. Such accumulation takes longer time at lower
write intensity as long as above condition is satisfied. This long
time delayed stimulated Stokes in P-cell suggests that the decaying
rate of the amplitude of the spin wave is much slower than that in a
regular cell, because the paraffin coating helps in protecting the
prepared spin wave from being destroyed by the collision with the
walls, thus making the long lifetime possible. This long lifetime is
a necessary condition to achieve long coherence time of the spin
wave.
\par
\begin{figure}[htbp]
\includegraphics[width=8.6cm,height=2.9cm]{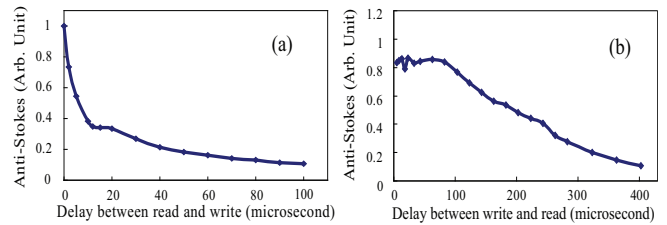}
\caption{Decay of Anti-Stokes V.S. delay between read and write.
(a) The decay of anti-Stokes signal at $\theta = 2^{o}$. The
diameter of the write laser is 3mm and the read is 10mm. Decay
time in paraffin-coated cell is composed of fast part, which is
10$\mu$s, and slow part, which is 80$\mu$s. (b)The decay of
anti-Stokes at $\theta = 0^{o}$. The diameter of the write laser
is 6mm and the read is 10mm. It only shows a slow decay with time
of 320$\mu$s.} \label{fig:fig4}
\end{figure}

Furthermore the coherence time of the spin wave generated in the
writing process can be directly measured by the reading process,
this is the T-CARS method as described before. The write pulse is
shut off at the peak of the stimulated Stokes, which is 40$\mu$s
for 5mW write power. The intensity of the anti-Stokes signal
during the retrieval process is recorded and plotted against the
time delay between the read and write pulses. The decay of the
anti-Stokes signal is caused by decoherence of the spin wave.
Fig~\ref{fig:fig4}(a) (write power 5mW, write beam dia.3 mm, read
power 80mW, read beam dia.10mm) shows the entire decay of the
anti-Stroke signal (relative intensity with respect to 0 delay
between read and write) in a P-cell, which displays two
exponential decays. Below 20$\mu$s is a fast decay with
characterized time 10$\mu$s. The later part is a slow decay,
characterized time about 80$\mu$s. This long decay time is in
agreement with the results from the long-delayed stimulated Raman
and suggested this long decay is resulting from the long lifetimes
of the spins due to reduced inelastic collision with the paraffin
coated surface. We still need to address the mechanism causing the
initial fast decay in Fig~\ref{fig:fig4}(a).
\par
\par
One reason for the fast decay is caused by the diffusion of the spin
wave out of the effective retrieval region covered by the reading
laser, as suggested by M. D.Eisaman \textit{et al.}~\cite{8}, but
this is not the major one because further studies at different write
beam widths and at flatter read power distribution covering the
entire cell shows the fast decay is always there under $\theta =
2^{o}$ detection configuration. As pointed out in the Duan
\textit{et al's} original paper~\cite{2} and investigated in detail
recently by Zhao \textit{et al}~\cite{18} in cold atom ensemble,
there is a dephasing of spin wave induced by random motion of the
atoms and can be easily understood in the case of single excitation.
\par
Upon detection of the forward Stokes signal, a singly excited spin
wave is created in the atomic ensemble in the form of
\begin{equation}
\notag |\psi_{spin}\rangle = \frac{1}{\sqrt{N}}\sum e^{i\triangle
k\cdot r_{j}} |1...2_{j}...1\rangle
\end{equation}
with $\triangle k = k_w - k_s$ the wave vector of the spin wave,
where $k_w$,$k_s$ are those of the write laser and the Stokes. The
atoms position changes to $r_{j}(t) = r_{j} + \triangle r{t}$ at
later time and will introduce a random phase shift. This will
cause a drop in the overlap between the spin waves $\langle
\psi_{spin}(t)|\psi_{spin}(0)\rangle$ , and thus the decay in
retrieval efficiency in the reading process.
\par
Under condition $\triangle k\cdot l << \pi $, with l the dimension
of the atomic ensemble, such dephasing is negligible, but this is
not the case in our skewed detection configuration. With $\theta =
2^{o}$, and $|\triangle k| = |k_w - k_s| \approx k_{w}sin\theta$,
then $\lambda_{spin} \approx \lambda_{w}/sin\theta = 23\mu m $.
This is much smaller than the size (and the maximum value of
$\triangle r$ ) of our cell, and the dephasing caused by the
random motion will be severe and the dominating factor in this
case. In order to minimize the effect of dephasing, the created
spin wave has to have a longer wavelength. The longest wavelength
of the spin wave is achieved under collinear detection
configuration with $\theta = 0^{o}$, where $\lambda_{spin} \approx
4.4$cm. In fact to have the wavelength of the spin wave on the
order of cm comparable to the dimension of the sample cell, the
angle $\theta$ between the write and Stokes signal has to be
smaller than $10^{-4}$ arc. So the only meaningful configuration
to minimize the dephasing in our experiment is the collinear one.
\par
Though the above argument is applied to the single excitation, it
can be extended to the multi-excitation case and provides the basis
for the improvement in the stimulated Raman region. A T-CARS
experiment was carried out under the collinear configuration and the
anti-Stokes decay is shown in Fig~\ref{fig:fig4}(b), the initial
fast decay due to the dephasing by thermal motion disappears, only a
long decay remains and the fitted ($e^{-t^{2}/\tau^{2}}$)
exponential decay time is $\tau \sim$320 $\mu$s. This demonstrated
the robust of the long wavelength of the spin wave against the
dephasing and the dominant decoherence effect here becomes the spin
lifetime. This result in the thermal vapor cell agrees with the
recent study by Zhao \textit{et al} in cold atom system~\cite{18}.
\par
In conclusion, we demonstrate the prolonged coherence time
$\tau\sim$320 $\mu$s of the collective spin wave in a paraffin
coated $^{87}$Rb cell. The coherence time is one hundred times
longer than the previous reported in the thermal vapor system. In
such system, the two major factors dominating the decoherence
process are the lifetime of the spin and the dephasing of the spin
wave due to thermal motion. The paraffin coating helps prolong the
lifetime of the spin by reducing the inelastic collision with the
walls which is manifested by the long delayed stimulated Stokes
signal under weak write laser, and the long decay of anti-Stokes
signal with T-CARS method. The dephasing can be overcome by choosing
the correct detection configuration and preparing the long
wavelength spin wave, this is manifested in the disappearing of the
fast decay in the anti-Stokes measurement. This coherence time
though measured in the stimulated region, offers at least an upper
limit for memory time in quantum repeater. In our study we also
demonstrated that the buffer-gas-free cell also greatly suppresses
the collision-induced fluorescence background.
\par
We gratefully thank Prof. J\"oerg. Schmiedmayer for discussions.
Dr. ChengZhi Peng, Dr. Dong Yang and ChengXi Yang offered great
help in the designing and programming of the data collecting
system. This work was supported by the National Natural Science
Foundation of China under Grant No.10474053 and No.10574162 and
Tsinghua University 985 Grant/051110001.

\end{document}